%% file: main.tex
\documentclass[a4paper]{easychair}
\usepackage[utf8]{inputenc}
\usepackage[T1]{fontenc}
\usepackage{amsmath}
\usepackage{amssymb}
\usepackage{wrapfig}
\usepackage{hyperref}
\usepackage{xspace}
\usepackage[shortlabels]{enumitem}
\usepackage{pifont}
\usepackage{longtable}
\usepackage{tabularray}

\theoremstyle{definition}
\newtheorem{example}{\textbf{Example}}{\itshape}{\rmfamily}

\usepackage{todonotes}
\usepackage{enumitem}  
\setlist{nolistsep} 

\usepackage{placeins}

\newcommand*{\anddot}{%
	\mathclose{}%
	\nonscript\mskip.5\thinmuskip
	\boldsymbol{.}%
	\;%
	\mathopen{}%
}

\newcommand{\quicksort}{\texttt{Quicksort}}
\newcommand{\mergesort}{\texttt{Mergesort}}
\newcommand{\insort}{\texttt{Insertionsort}}

\usepackage{while}

\newcommand{\pv}[1]{\texttt{#1}}

\usepackage{ stmaryrd }

\newcommand{\natsort}{\mathbb{N}}

\newcommand{\msf}[1]{\ensuremath{\mathsf{#1}}\xspace}
\newcommand{\Ind}{\msf{Ind}}
\newcommand{\cnf}{\msf{cnf}}
\newcommand{\nil}{\msf{nil}}
\newcommand{\cons}{\msf{cons}}

\usepackage{bussproofs}

\newcommand{\vampire}{\textsc{Vampire}}

\usepackage{url}

\usepackage{rotating}
\usepackage{multirow}

\usepackage{subcaption}

\usepackage{float}
\newfloat{lstfloat}{htbp}{lop}
\floatname{lstfloat}{Listing}

\newcommand{\CommentedOut}[1]{}
\newcommand{\ToCheck}[1]{{\color{blue}#1}}

\renewcommand{\paragraph}[1]{\par\noindent{{\bf #1}}}

\begin{document}

\titlerunning{Saturating Sorting without Sorts}
\title{
	Saturating Sorting without Sorts
}
\author{
  Pamina Georgiou
   \and
  M\'arton Hajdu
  \and
  Laura Kov{\'a}cs
}
\authorrunning{Georgiou, Hajdu and Kov{\'a}cs}
\institute{
  TU Wien, Austria
}
\maketitle        

\input{abstract.tex}

\input{intro.tex}

\input{preliminaries.tex}

\input{semantics.tex}

\input{motivating.tex}

\input{induction.tex}


\input{experiments.tex}

\input{related.tex}

\input{conclusion.tex}

\smallskip

\paragraph{Acknowledgements.} This research was supported by  the ERC Consolidator Grant ARTIST 101002685, the Austrian FWF SFB project SpyCoDe F8504, and the SecInt Doctoral College funded by TU Wien.

\bibliographystyle{splncs04}
\bibliography{bib.bib}

\input{appendix.tex}

\end{document}

%% file: abstract.tex

\begin{abstract}
  We present a first-order theorem proving framework  for establishing the correctness of functional programs implementing sorting algorithms with recursive data structures. 
  We formalize the semantics of recursive programs in many-sorted first-order logic and integrate sortedness/permutation properties within our first-order formalization. Rather than focusing on sorting lists of elements of specific first-order theories, such as integer arithmetic, our list formalization relies on a sort parameter abstracting (arithmetic) theories and hence concrete sorts. We formalize the permutation property of lists in first-order logic so that we automatically prove verification conditions of such algorithms purely by superpositon-based first-order reasoning.  Doing so, we adjust recent efforts for
  automating inducion in saturation. We advocate a compositional approach for automating proofs by induction  required to verify functional programs implementing and preserving sorting and permutation properties over parameterized list structures. Our work turns saturation-based first-order theorem proving into an automated verification engine by (i) guiding automated inductive reasoning with manual proof splits and (ii) fully automating inductive reasoning in saturation.  We showcase the applicability of our framework over recursive sorting algorithms, including Mergesort and Quicksort. 
%
%
%
\end{abstract}

%% file: intro.tex

\section{Introduction}
Sorting algorithms are integrated parts of any modern programming language, hence ubiquitous in computing, which naturally triggers the demand of validating the functional correctness of sorting routines. Such algorithms typically 
  implement recursive/iterative operations over potentially unbounded data structures, for instance lists or arrays, combined with
  arithmetic manipulations of numeric data types, such as naturals, integers or reals.  
  Automating the formal verification of sorting routines therefore brings the challenge of automating recursive/inductive reasoning in extensions and combinations of first-order theories,
   while also addressing the reasoning burden arising from design choices made for the purpose of efficient sorting. Most notably,  \quicksort{}~\cite{hoare1962quicksort} is known to be easily implemented when making use of recursive function calls, for example,
  as given in Figure~\ref{fig:quicksort},  whereas purely procedural implementations of \quicksort{} require additional recursive data structures such as stacks. While \quicksort{} and other sorting routines have been proven correct by means of manual efforts~\cite{foley1971proof},  proof assistants~\cite{nipkow2021functional, walther2004verification, beckert2017proving},  abstract interpreters~\cite{DBLP:conf/popl/GulwaniMT08}, or model checkers~\cite{DBLP:conf/cav/JhalaM07},
to the best of our knowledge such correctness proofs so far have not been fully automated with saturation-based automated reasoning. 

\begin{figure}[t]
\begin{minipage}[]{1\textwidth}%
			\begin{lstlisting}[mathescape]
datatype  a' list = $\nil$ | $\cons$(a', (a' list))

quicksort :: a' list $\rightarrow$ a' list
quicksort($\nil$) = $\nil$
quicksort($\cons\pv{(x, xs)) = }$
	append(
		quicksort(filter$_<$(x, xs)) , 
		$\cons\pv{(x, }$quicksort(filter$_\geq$(x, xs))))

append :: a' list $\rightarrow$ a' list $\rightarrow$ a' list
append($\nil$, xs) = xs
append($\cons$(x, xs), ys) = $\cons$(x, append(xs, ys))
\end{lstlisting} 
\caption{Recursive algorithm of \quicksort{}, using the recursive function definitions \pv{append}, $\pv{filter}_<$ and $\pv{filter}_{\geq}$ over lists of sort  $a$. 
}
			\label{fig:quicksort}
		\end{minipage}\vspace*{-\baselineskip}
\end{figure}

\emph{In this paper we aim to verify the partial correctness of
	functional programs with recursive data structures, in an automated manner by using saturation-based first-order theorem proving.}
      To achieve this, we turn the automated first-order reasoner into a complementary approach to interactive proof assistance: (i) we rely on manual guidance in splitting inductive proof goals into subgoals (Sections~\ref{sec:quicksort} and \ref{sec:lemmas}), but (ii) fully automate inductive proofs in saturation-based reasoning (Section~\ref{sec:compind}). The crux of our approach is a compositional reasoning setting based on
superposition-based first-order theorem proving~\cite{kovacs2013first} with native support for induction~\cite{hajdu2022getting}  and
 first-order theories of recursively defined data types~\cite{kovacs2017coming}. 
 We extend this setting to support the first-order theory of list data structures parameterized by an abstract background theory/sort $a$ and advocate  \textit{computation induction} for induction on recursive function calls. 
 As such, our framework allows us to automatically discharge manually split verification conditions that require inductive proofs, without requiring manually proven or a priori given inductive annotations such as loop invariants, nor user input to perform proofs by induction. 
Doing so, we automatically derive induction axioms during \textit{saturation} to establish the functional correctness of the recursive
implementation of \quicksort{} from Figure~\ref{fig:quicksort} by means of automated first-order reasoning. 
In a nutshell, we proceed as follows.  \smallskip

%


        (i) We formalize the \emph{semantics of functional programs} in extensions of the first-order theory of lists (Section~\ref{sec:semantics}), allowing us to quantify over lists.
        Rather than focusing on lists with a specific background theory, such as integers/naturals,
        our formalization relies on a parameterized sort/type $a$ abstracting specific (arithmetic) theories. To this end, we impose that the sort $a$ has a linear order $\leq$. 
        Doing so, we remark that one of the major reasoning burdens towards establishing the correctness of sorting algorithms comes with formalizing permutation properties, for example that two lists are permutations of each other. Universally quantifying over permutations of lists is, however, not a first-order property and hence reasoning about list permutation requires higher-order logic. While counting and comparing the number of list elements is a viable option to formalize permutation equivalence in first-order logic, the necessary arithmetic reasoning adds an additional burden to the underlying prover. We overcome this challenge by introducing an effective first-order formalization of 
        \emph{permutation equivalence} over parameterized lists. Our permutation equivalence property encodes \textit{multiset} operations over lists, eliminating the need of counting list elements, and therefore arithmetic reasoning, or fully axiomatizing (higher-order) permutations. 
\smallskip

(ii) We revise inductive reasoning in first-order theorem proving (Section~\ref{sec:compind}) and introduce \textit{computation induction} as a means to tackle \textit{divide-and-conquer} algorithms. We, therefore, extend the first-order reasoner with an inductive inference based on the \textit{computation induction schema} and outline its necessity for recursive sorting routines.\smallskip

(iii) We leverage \emph{first-order theorem proving for compositional proofs} of recursive parameterized sorting algorithms (Section \ref{sec:quicksort}),  in particular of \quicksort{} from Figure~\ref{fig:quicksort}. By embedding the application of induction directly in saturation-based proving, we automatically discharge manually split proof obligations. Each such condition represents a first-order lemma, and hence a proof step. We emphasize that
the only manual effort in our framework comes with splitting formulas into multiple lemmas (Section~\ref{sec:lemmas:split});  each lemma is established automatically by means of automated theorem proving with built-in induction.
That is, all our lemmas/verification conditions are automatically proven by means of  structural and/or computation induction during the saturation process. 
We do not rely on user-provided inductive properties, nor on user guidance to perform proofs by induction, but 
generate inductive hypotheses/invariants via inductive inferences automatically as logical consequences 
of our program semantics. \smallskip

(iv) 	 We note that sorting algorithms often follow a divide-and-conquer approach (see Figure~\ref{fig:div-and-conq}).
 We, thus, apply our approach on other sorting routines and investigate a generalized set of manual proof splits/lemmas that is applicable to verify functional sorting algorithms on recursive data structures (Section~\ref{sec:lemmas}). 
\smallskip 
%

(v) We demonstrate our findings (Section~\ref{sec:experiments})  by implementing our approach on top of the \vampire{} theorem prover~\cite{kovacs2013first}, providing thus a fully automated tool support towards validating the functional correctness of sorting algorithms.

%% file: preliminaries.tex
\section{Preliminaries} \label{sec:prelim}

We assume familiarity with standard first-order logic (FOL) and briefly
introduce saturation-based proof search in first-order theorem
proving~\cite{kovacs2013first}. 
\noindent\paragraph{Saturation.} Rather than using arbitrary first-order formulae $G$, most first-order theorem provers rely on a clausal representation $C$
of $G$. The task of first-order theorem proving is to establish  that a
formula/goal $G$ is a logical consequence of a set $\mathcal{A}$ of
clauses, including assumptions. Doing so, first-order provers clausify
the negation  $\neg G$ of $G$ and derive that the set  $S =
\mathcal{A}\cup\{\neg G\}$ is unsatisfiable\footnote{for simplicity, we
 denote by  $\neg G$ the clausified form of the negation of $G$}. To
this end, first-order
provers {\it saturate} $S$ by computing all logical consequences of $S$ with respect to some sound inference
system $\mathcal{I}$. A sound inference system $\mathcal{I}$ derives a
clause $D$ from clauses ${C}$ such that
$C\rightarrow D$.
The saturated set of $S$ w.r.t.
$\mathcal{I}$ is called the {\it closure} of $S$ w.r.t. $\mathcal{I}$,
whereas the process of deriving the closure of $S$ is called
\textit{saturation}.
By soundness of $\mathcal{I}$, if the closure of $S$ contains the empty clause $\square$, the original
set $S$ of clauses is unsatisfiable, implying the validity of
$\mathcal{A}\rightarrow G$; in this case, we established a
\emph{refutation} of $\neg G$ from $\mathcal{A}$, hence a proof of
validity of $G$.

The \textit{superposition calculus} is a common inference
system used by saturation-based provers for FOL with equality~\cite{robinson2001handbook}.
The superposition calculus is {sound} and \textit{refutationally
  complete}:
for any unsatisfiable formula $\neg G$, superposition-based saturation
derives the empty clause $\square$ as a logical consequence of $\neg G$.

\noindent\paragraph{Parameterized Lists.} We use the first-order theory of
recursively defined datatypes~\cite{kovacs2017coming}. In particular,
we consider the list datatype with two constructors \nil{} and
$\cons(x, xs)$, where \nil{} is the empty list and $x$ and $xs$
are respectively the head and tail of a list.  
We introduce a type parameter $a$ that abstracts the
sort/background theory of the list elements. Here, we impose the
restriction that the sort $a$ 
has a linear order $<$, that is, a binary relation which is reflexive, antisymmetric, transitive and total. For simplicity, we also use $\geq$ and $\leq$
as the standard ordering extensions of $<$. 
We write $xs_a,
ys_a,zs_a$ to mean that the lists $xs,ys,zs$ are parameterized by sort
$a$; that is their elements are of sort $a$. Similarly, we use $x_a, y_a,z_a$ to mean that the list elements
$x,y,z$ are of sort $a$.  Whenever it is clear from the context, we omit specifying the sort $a$. 

	\noindent\paragraph{Function definitions.}  We make the
        following abuse of
        notation. For some function \pv{f} in some program \pv{P}, we use the notation \pv{f(arg$_1$, ...)} to refer to function definitions/calls appearing in the input algorithm, while the mathematical notation $f(arg_1, ...)$ refers to its counterpart in our logical representation as per our first-order semantics introduced in Section \ref{sec:semantics}.

%% file: semantics.tex

\section{First-Order Semantics of Functional Sorting Algorithms} \label{sec:semantics}
We outline our formalization of recursive sorting algorithms in the full
first-order theory of parameterized lists.
%
\subsection{Recursive Functions in First-Order Logic}\label{sec:semantics:divide}
We investigate recursive algorithms written in a functional coding
style and defined over lists using list constructors.
That is, we consider recursive functions \pv{f} that manipulate the
empty list $\nil$ and/or the list $\cons(x, xs)$. 

Many recursive sorting algorithms, as well as other recursive
operations over lists, 
implement 
a \emph{divide-and-conquer} approach: let \pv{f} be a function following such a pattern, \pv{f} uses (i) a \textit{partition
  function} to divide 
 $list_a$, that is a $list$ of sort $a$, into two smaller sublists upon which  \pv{f}
is recursively applied to, and 
(ii) calls a \textit{combination function} that puts together the result
of the recursive calls of \pv{f}.
Figure \ref{fig:div-and-conq} shows such a divide-and-conquer pattern,
where  the  partition function \pv{partition} uses an invertible operator
$\circ$, with $\circ^{-1}$ being 
the complement of $\circ$; \pv{f} is applied to the results of
$\circ$  before these results are merged using the combination
function \pv{combine}.

\noindent\begin{wrapfigure}{r}{0.5\textwidth}
	\vspace*{-1em}
	{\small 	
			\begin{lstlisting}[mathescape]
f ::  a' list $\rightarrow ... \rightarrow$ a' list
f($\nil, ...$) = $\nil$
f($\cons$($\pv{y}, \pv{ys}), ...$)= combine(
	f(partition$_{\circ}$($\cons$($\pv{y}, \pv{ys}$))),
	f(partition$_{\circ^{-1}}$($\cons$($\pv{y}, \pv{ys}$))))
			\end{lstlisting} 
			\caption{Recursive divide-and-conquer approach.}
			\label{fig:div-and-conq}
		}
	\end{wrapfigure}

Note that the recursive function \pv{f} of Figure
\ref{fig:div-and-conq} is 
defined via the declaration $f :: a' list
\rightarrow ... \rightarrow a' list$, where $...$ denotes further
input parameters. 
We formalize the first-order semantics of \pv{f}  via the function
$f \colon (list_a \times ...) \mapsto list_a $, by translating the 
inductive function definitions \pv{f} to the following 
first-order formulas with parameterized lists  (in first-order logic, function
definitions can be considered as universally quantified equalities): 

\begin{align}\label{eq:div-and-conq}
	\begin{array}{rcll}
	f(\nil)  & =& \nil & \\
	\forall x_a, xs_a \anddot f(\cons(x, xs)) &=&combine(
        f(partition_{\circ}(\cons(x, xs))),  \\
	&&f(partition_{\circ^{-1}}(\cons(x, xs)))
	).
\end{array}
\end{align}
The recursive divide-and-conquer pattern of
Figure~\ref{fig:div-and-conq}, together with the first-order
semantics~\eqref{eq:div-and-conq} of \pv{f}, is used in
Sections~\ref{sec:quicksort}-\ref{sec:lemmas} for proving correctness
of the \quicksort{} algorithm (and other sorting algorithms), as well
as for applying lemma generalizations for divide-and-conquer list
operations.
We next introduce
our first-order formalization for specifying that \pv{f} implements a
sorting routine. 

\subsection{First-Order Specification of
  Sorting Algorithms}\label{sec:algorithms:sort}\label{sec:algorithms:perm}
We consider a specific function  instance of \pv{f} implementing a
sorting algorithm,  expressed through $sort :: a' list
\rightarrow a' list$. The functional behavior of $sort$  needs to
satisfy two specifications implying the functional correctness of
$sort$: (i) sortedness and
(ii) permutations  equivalence of the list computed
by $sort$. 

\noindent\paragraph{(i) Sortedness:}  \emph{The list computed
by the $sort$ function must be sorted w.r.t. some linear order $\leq$ over the type $a$  of
list elements.} We define a parameterized version of this sortedness property 
using an inductive predicate $sorted$ as follows: 
\begin{equation}\label{eq:elem_list_sorted}
  \begin{array}{rl}
		sorted(\nil) & = \top \\
		\forall x_a, xs_a \anddot sorted(\cons(x, xs)) & =
                                                                 (elem_\leq
                                                                 list(x,xs)
                                                                 \wedge
                                                                 sorted(xs)), 
  \end{array}
  \end{equation}
        where $elem_\leq list(x,xs)$ specifies that $x\leq y$ for any
        element $y$ in $xs$. 
Proving correctness of a sorting algorithm $sort$
      thus  reduces to proving the validity of: 
	\begin{align}\label{prop:sorted}
		\forall xs_a \anddot sorted(sort(xs)).
	\end{align}


\noindent\paragraph{(ii) Permutation Equivalence:} \emph{The list computed
by the  $sort$ function is  a permutation of the input list to the
$sort$ function}. In other
words the input and output lists of $sort$ are permutations of each
other, in short permutation equivalent.

Axiomatizing
permutations requires quantification over relations and is
thus not expressible in first-order logic~\cite{laneve1996axiomatizing}. 
A common approach to prove permutation equivalence of two lists 
is to count the occurrences of each element in both lists respectively and
compare these numbers per list element. Yet, counting adds a burden of arithmetic reasoning over naturals to the underlying prover, calling for additional applications of
mathematical induction. 
We overcome these challenges of expressing permutation equivalence as follows. We
introduce a family of functions $filter_Q$ manipulating lists, with  the
function  $filter_Q$ being parameterized by a predicate $Q$ and as
given in Figure \ref{fig:filter_ms}.

\begin{wrapfigure}{l}{0.65\textwidth}
{\small 	
\begin{lstlisting}[mathescape]
filter$_{Q}$ :: a' $\rightarrow$ a' list $\rightarrow$ a' list
filter$_{Q}$($\pv{x}, \nil$) = $\nil$
filter$_{Q}$($\pv{x}, \cons$($\pv{y}, \pv{ys}$))=
	if (Q($\pv{y},\pv{x}$)) {
		$\cons$($\pv{y}, $filter$_{Q}$($\pv{x}, \pv{ys}$))
	} else {
		filter$_{Q}(\pv{x}, \pv{ys})$
	}
\end{lstlisting} \vspace{-1em}
\caption{Functions $filter_Q$ filtering elements of a list, by using a
  predicate $Q(y,x)$ over list elements $x,y$.}
\label{fig:filter_ms}
}
\end{wrapfigure}


%
%

In particular, given an element $x$ and a list $ys$, the functions $filter_=$, $filter_<$,
and $filter_\geq$ compute the maximal sublists of $ys$ that contain only equal, resp.
smaller and greater-or-equal elements to $x$. 
Analogously to counting the multiset multiplicity of $x$ in $ys$ via counting functions,
we compare lists given by $filter_=$, avoiding the need to count the number of occurrences of $x$ and hence prolific axiomatizations of arithmetic.
Thus, to prove that the input/output lists of $sort$  are permutation
equivalent, we show that, for every list element $x$, the
results 
of applying \pv{filter$_{=}$} to the input/output list of $sort$ are
the same over all elements. Formally, we have the following first-order property of
permutation equivalence: 
\begin{align}\label{prop:equal_msets}
	\forall x_a, xs_{a} \anddot filter_=(x, xs) &= filter_=(x, sort(xs)).
\end{align}

\section{Computation Induction in Saturation} \label{sec:compind}\label{sec:semantics:induction}
In this section, we describe our reasoning
  extension to saturation-based first-order theorem proving, in order
  to support inductive reasoning for recursive sorting algorithms as
  introduced in Section \ref{sec:semantics}. Our key reasoning
  ingredient comes with a structural
  induction schema of \textit{computation induction}, which we
  directly integrate in the saturation proving process. 
	\newline Inductive reasoning has recently been embedded in saturation-based
theorem proving~\cite{hajdu2022getting},
by extending the superposition calculus with a new inference rule
based on \emph{induction axioms}:
%
	
	\medskip
	\begin{tabular}{c p{.1\linewidth} p{.55\linewidth}}
		\multirow{2}{*}{
			\AxiomC{$\overline{L}[t] \lor C$}
			\LeftLabel{$(\Ind)$}
			\UnaryInfC{$\cnf(\neg F \lor C)$}
			\DisplayProof
		}
		&
		\multirow{2}{*}{where}
		&
		(1) $L[t]$ is a quantifier-free (ground) literal,\\
		& & (2) $F\rightarrow \forall x.L[x]$ is a valid \textit{induction
			axiom}, \\
		& & (3) $\cnf(\neg F \lor C)$ is the clausal form of $\neg F \lor C$. 
	\end{tabular}\bigskip

An \emph{induction axiom} refers to an instance of a valid
induction schema. In our work, we use 
structural and computational induction schemata.

In particular, we use the following \emph{structural induction} schema over
lists: 
%
\begin{equation}\label{eq:list-schema}
	\begin{pmatrix}L[\nil]\land\forall x,ys.(L[ys]\to L[\cons(x,ys)])\end{pmatrix}\to\forall zs.L[zs]
\end{equation}

Then, considering the induction axiom resulting from applying 
	schema~\eqref{eq:list-schema} to $L$, we obtain the following \Ind instance for lists:\vspace{-.5em}
	\begin{prooftree}
		\AxiomC{$\overline{L}[t] \lor C$}
		\UnaryInfC{$\begin{matrix}
				\overline{L}[\nil]\lor L[\sigma_{ys}] \lor C\\
				\overline{L}[\nil]\lor \overline{L}[\cons(\sigma_x,\sigma_{ys})] \lor C
			\end{matrix}$}
	\end{prooftree}
	where  $t$ is a ground term of sort list, $L[t]$ is ground, and
	$\sigma_x$ and $\sigma_{ys}$ are fresh constant symbols.
	The above \Ind instance yields two clauses as conclusions and is applied during the saturation process.\smallskip


Sorting algorithms, however, often require induction
axioms that are more complex than instances of structural
induction~\eqref{eq:list-schema}. Such axioms are typically instances of the
{computation/recursion induction} schema, arising from divide-and-conquer strategies as introduced in Section~\ref{sec:semantics:divide}.
Particularly, the complexity arises due to the two
  recursive calls on different parts of the original input list
  produced by the \textit{partition} function that have to be taken
  into account by the induction schema.
We therefore use the following \emph{computation induction}
schema over lists:
{\small 
	\begin{equation}\label{eq:quicksort-schema}
		\begin{pmatrix}L[\nil]\land\forall
			x,ys.\begin{pmatrix}\begin{pmatrix}L[partition_\circ
					(x,ys)]\land\\
					L[partition_{\circ^{-1}}(x,ys)]\end{pmatrix}\to
				L[\cons(x,ys)])\end{pmatrix}\end{pmatrix}\to\forall
		zs.L[zs]
\end{equation}}

yielding the following instance of \Ind that is applied during saturation: 
	
	\begin{prooftree}\label{eq:ind:comp}
          \hspace*{-1em}	\AxiomC{$\overline{L}[t] \lor C$}
		\UnaryInfC{$\begin{matrix}
				\overline{L}[\nil]\lor L[partition_\circ(\sigma_x,\sigma_{ys})] \lor C\\
				\overline{L}[\nil]\lor L[partition_{\circ^{-1}}(\sigma_x,\sigma_{ys})] \lor C\\
				\overline{L}[\nil]\lor \overline{L}[\cons(\sigma_x,\sigma_{ys})] \lor C
			\end{matrix}$}
	\end{prooftree}
	where  $t$ is a ground term of sort list, $L[t]$ is ground, $\sigma_x$
	and $\sigma_{ys}$ are fresh constant symbols, and $partition_\circ$ and
	its complement refer
	to the functions that partition lists into sublists within the actual sorting algorithms. 
%

\CommentedOut{
	\ToCheck{
		In what follows, upon performing an $\mathsf{Ind}$ inference, we say
		that the induction axiom $F\rightarrow \forall x.L[x]$ is
		\textit{applied on} the clause $\overline{L}[t]\lor C$, or
		alternatively speaking $\overline{L}[t]\lor C$ is \textit{inducted
			upon}. In addition, we also say that we 
		\textit{induct on term} $t$ or \textit{induct with respect to} $t$ in
		clause $\overline{L}[t]\lor C$ with induction axiom $F\rightarrow
		\forall x.L[t]$.
	}
}


%% file: motivating.tex

\section{Proving Recursive \quicksort} \label{sec:quicksort}

%


We now describe our approach towards proving
the correctness of the recursive parameterized version of
\quicksort{}, as given in Figure \ref{fig:quicksort}.
Note that \quicksort{} recursively sorts
two sublists that contain respectively smaller and
greater-or-equal elements than the pivot element $x$ of its input list.
We reduce the
task of proving the functional correctness of  \quicksort{} to the
task of proving the (i) sortedness property~\eqref{prop:sorted} and (ii) the permutation equivalence
property~\eqref{prop:equal_msets} of \quicksort{}.
As mentioned in 
Section~\ref{sec:algorithms:sort}, a similar reasoning is
needed for most sorting algorithms, which we evidence in
Sections~\ref{sec:lemmas}--\ref{sec:experiments}. 

\subsection{Proving Sortedness for \quicksort}\label{sec:prove:sorted}
Given an input list $xs$, we prove that \quicksort{} computes a
sorted list by
considering the property~\eqref{prop:sorted} instantiated for
\quicksort. That is, we prove: 
\begin{align}\label{prop:sortedqs}
	\forall xs_a \anddot sorted(quicksort(xs))
\end{align}
The sortedness property~\eqref{prop:sortedqs} of \quicksort{} is
proved via \emph{compositional reasoning}
over~\eqref{prop:sortedqs}. Namely, 
we enforce the following two properties that
together imply~\eqref{prop:sortedqs}:\\[-.5em]
\begin{enumerate}[label={{\bf (S\arabic*)}},wide=0em,leftmargin=.5em]
  \item\label{S1} By using the linear order $\leq$ of the background theory $a$,
    for any element $y$ in the sorted list $quicksort(filter_<(x,xs))$
    and any element $z$ in the sorted list $quicksort(filter_\geq(x,xs))$,
    we have $y\leq x\leq z$.\\[-.5em]
  \item \label{S2}
The functions $filter_<$ and $filter_\geq$
    of Figure~\ref{fig:filter_ms} are correct. That is, filtering
    elements from a list that are smaller, respectively  greater-or-equal,
    than 
    an element $x$ results in sublists only containing elements
    smaller than, respectively greater-or-equal, than $x$.\\[-.5em]
\end{enumerate}

    Similarly to~\eqref{eq:elem_list_sorted} and to express
    property~\ref{S2}, we introduce the inductively defined predicates $elem_\leq list:: a' \to a'list \to bool$ and $list_\leq list:: a' list \to a' list\to bool$:
\begin{equation} \label{eq:list_ge_elem}
	\begin{split}
	\forall x_a \anddot elem_\leq list(x,\nil) & = \top\\
	\forall x_a, y_a, ys_a \anddot elem_\leq list(x,\cons(y,ys)) & = x \leq y \wedge elem_\leq list(x,ys),
	\end{split}
      \end{equation}
 and 
\begin{equation} \label{eq:list_ge_list}
	\begin{split}
	\forall ys_a \anddot list_\leq list(\nil,ys) & = \top\\
	\forall  x_a, xs_a, ys_a \anddot list_\leq list(\cons(x,
          xs),ys) & = (elem_\leq list(x,ys) \wedge list_\leq list(xs, ys)).
	\end{split}
\end{equation}
Thus, for some element $x$ and lists $xs$, $ys$, 
  we express that $x$ is smaller than or equal to any element of $xs$ by
  $elem_\leq list(x,xs)$. Similarly,  $ list_\leq list(xs, ys)$
states that every element in list $xs$ is smaller than or equal to any element in $ys$. 
 
The inductively defined predicates of \eqref{eq:list_ge_elem}--\eqref{eq:list_ge_list} allow
us to express necessary lemmas over list operations preserving the
sortedness property~\eqref{prop:sortedqs}, for example, to prove that
appending sorted lists yields a sorted list.
\CommentedOut{
Note that the $<$ and $\geq$ relations and their axiomatizations also depend on the type parameter $a$. That is,  $<$ is essentially an uninterpreted function such that  $\geq$ is, logically speaking, simply its negation. However, we consider a small axiomatization of $<$ as a linear order outlined in Section \ref{sec:prelim}.  Our proof is, thus, more general than standard variations considering specific background theories. Specifically, in terms of state space this means that we over-approximate all possible states that might not need to be proved correct in a specific sort, say integer theory. 
}


Proving properties \ref{S1}--\ref{S2}, and
hence deriving the sortedness property~\eqref{prop:sortedqs} of
\quicksort, requires \emph{three first-order lemmas}
in addition to the first-order
semantics~\eqref{eq:div-and-conq} of \quicksort.
Each of these lemmas is automatically proven by  saturation-based
theorem proving using the structural and/or computation induction
schemata of~\eqref{eq:list-schema} and~\eqref{eq:quicksort-schema};
hence, by compositionality, we obtain \ref{S1}--\ref{S2}
implying~\eqref{prop:sortedqs}.  We
next discuss these three lemmas and  outline the complete (compositional) proof of the sortedness property~\eqref{prop:sortedqs} of
\quicksort. 

\begin{enumerate}[label={{$\bullet$}},wide=0em,leftmargin=0em]
\item
  In support of~\ref{S1}, 
lemma \eqref{prop:qs:lemma1} expresses that for two \emph{sorted} lists $xs,
ys$ and a list element $x$,
such that $elem_{\leq}list(x,xs)$ holds
and all elements of the constructed list  $\cons(x, xs)$ are
greater than or equal to all elements in $ys$,
the result of concatenating $ys$ and $\cons(x, xs)$ yields a sorted list. Formally, we have 
\begin{equation} \label{prop:qs:lemma1}
	\begin{array}{ll}
		\forall x_a, xs_a, ys_a \anddot & 
			\big( sorted(xs) \wedge sorted(ys) \wedge 
			elem_\leq list(x,xs)  \wedge \\
			& ~list_\leq list(ys,\cons(x, xs)) \big) \\
		&	\rightarrow
			sorted(append(ys,\cons(x, xs)))
	\end{array}
      \end{equation}
  \item In support of~\ref{S2}, 
we need to establish that filtering greater-or-equal elements for some
list element $x$ results in a list whose elements are greater-or-equal
than $x$. In other words, the inductive predicate  of \eqref{eq:list_ge_elem} is
invariant over sorting and filtering operations over lists.
\begin{equation} \label{prop:qs:lemma2}
	\begin{array}{l}
		\forall x_a, xs_a \anddot 
		elem_\leq list(x, {quicksort}(filter_\geq(x,xs))).
	\end{array}
      \end{equation}

    \item Lastly and in further support of \ref{S1}--\ref{S2},
      we establish that all elements of a list $xs$ are ``covered'' with
      the filtering
      operations \pv{filter$_\geq$} and \pv{filter$_<$} w.r.t. a list
      element $x$ of $xs$. Intuitively, a call of
      \pv{filter$_<$(x,xs)} results in a list containing all elements of $xs$ that are
      smaller than $x$, while  the remaining elements of $xs$ are
      those that are greater-or-equal than $x$ and hence are contained
      in $\cons(x,filter_{\geq}(x,xs))$. By applying \quicksort{}
        over the input list $xs$, we get: 
\begin{equation} \label{prop:qs:lemma3}
	\begin{array}{l}
		\forall x_a, xs_a \anddot	list_\leq list(
							 quicksort(filter_<(x,xs)), 
							 \cons(x, quicksort (filter_\geq(x,xs)))).
	\end{array}
\end{equation}
\end{enumerate}

The first-order lemmas~\eqref{prop:qs:lemma1}--\eqref{prop:qs:lemma3}
guide saturation-based proving to instantiate structural/computation
induction schemata and automatically 
derive the following induction axiom necessary to
prove~\ref{S1}--\ref{S2}, and hence sortedness of \quicksort: 
{\small
  \begin{equation}\label{prop:sorted:ih}
  \begin{array}{l}
	\Big( 
  sorted(quicksort(\nil)) \wedge \\
~~	\forall x_a, xs_a \anddot 
   \Big(
\begin{array}{l}
  sorted(quicksort(filter_\geq(x, xs))) \wedge\\
  sorted(quicksort(filter_<(x, xs)))
  \end{array}\Big) \rightarrow  sorted(quicksort (\cons(x, xs))  \Big) \\[.5em]
	\rightarrow
	\forall xs_a \anddot sorted(quicksort (xs)),
  \end{array}
\end{equation}
}

\noindent where axiom~\eqref{prop:sorted:ih}
is automatically obtained during saturation from the computation induction
schema~\eqref{eq:quicksort-schema}. Intuitively, the prover replaces $F$ by $sorted(quicksort(t))$ for some term $t$,  and uses $filter_<$ and $filter_\geq$ as $partition_\circ$ and $partition_{\circ^{-1}}$ respectively to find the necessary computation induction schema. We emphasize that this step is fully automated during the saturation run. 

The first-order
lemmas~\eqref{prop:qs:lemma1}--\eqref{prop:qs:lemma3}, together with
the induction axiom~\eqref{prop:sorted:ih} and the first-order
semantics~\eqref{eq:div-and-conq} of \quicksort, imply the sortedness
property~\eqref{prop:equal_msets} of \quicksort; this proof can
automatically be derived using saturation-based reasoning. Yet, the
obtained proof assumes the validity of each of
the lemmas~\eqref{prop:qs:lemma1}--\eqref{prop:qs:lemma3}. To 
eliminate this assumption, we propose to also prove
lemmas~\eqref{prop:qs:lemma1}--\eqref{prop:qs:lemma3} via
saturation-based reasoning. Yet, while lemma~\eqref{prop:qs:lemma1} is
established by saturation with structural
induction~\eqref{eq:list-schema} over lists, proving 
lemmas~\eqref{prop:qs:lemma2}--\eqref{prop:qs:lemma3} requires further
first-order formulas. In particular, for proving
lemmas~\eqref{prop:qs:lemma2}--\eqref{prop:qs:lemma3} via saturation,
we use four further lemmas, as follows.
\begin{enumerate}[label={{$\bullet$}},wide=0em,leftmargin=0em]
\item Lemmas~\eqref{prop:qs:lemma4}--\eqref{prop:qs:lemma5} indicate
  that the order of $elem_\leq list$ and $list_\leq list$ is preserved
  under $quicksort$, respectively. That is, 
\begin{align} \label{prop:qs:lemma4}
	\begin{split}
		\forall x_a, xs_a \anddot 
			elem_\leq list(x, xs) \rightarrow elem_\leq list(x,quicksort(xs))
	\end{split}
\end{align}
and 
\begin{align} \label{prop:qs:lemma5}
	\begin{split}
		\forall xs_a, ys_a \anddot 
		list_\leq list(ys, xs) \rightarrow list_\leq list(quicksort(ys),xs).
	\end{split}
\end{align}
\item 
Proving lemmas~\eqref{prop:qs:lemma4}--\eqref{prop:qs:lemma5},
however, requires two further lemmas that follow from saturation with
built-in computation and structural induction, respectively. Namely,
lemmas~\eqref{prop:qs:lemma6}--\eqref{prop:qs:lemma7} establish that
$elem_\leq list$ and $list_\leq list$ are also invariant over
appending lists. That is, 
\begin{equation} \label{prop:qs:lemma6}
	\begin{split}
		\forall x_a, y_a, xs_a, ys_a \anddot & \big(  y \leq x
                                                       \wedge
                                                       elem_\leq
                                                       list(y, xs)
                                                       \wedge
                                                       elem_\leq
                                                       list(y, ys)  
		\big) \\
		 & \rightarrow elem_\leq list(y, append(\cons(x, ys),xs))
	\end{split}
\end{equation}
and
\begin{equation} \label{prop:qs:lemma7}
	\begin{split}
		\forall xs_a, ys_a, zs_a \anddot & 
			\big( list_\leq list(ys, xs) \wedge list_\leq
                                                   list(zs, xs) \big) \\
		&\rightarrow list_\leq list(append(ys, zs), xs)
	\end{split}
\end{equation}
\end{enumerate}

With lemmas~\eqref{prop:qs:lemma4}--\eqref{prop:qs:lemma7}, we
automatically prove 
lemmas~\eqref{prop:qs:lemma1}--\eqref{prop:qs:lemma3} via
saturation-based reasoning. 
The complete automation of proving properties \ref{S1}--\ref{S2}, and
hence deriving the sortedness property~\eqref{prop:sortedqs} of
\quicksort{} in a compositional manner,
requires thus \emph{altogether seven lemmas} in addition to the first-order
semantics~\eqref{eq:div-and-conq} of \quicksort. \emph{Each of these lemmas
is automatically established via saturation with  built-in
induction. }
Hence, unlike interactive theorem proving, 
compositional proving with first-order theorem provers can be
leveraged to eliminate the need to a priori specifying necessary
induction axioms. 


\subsection{Proving Permutation Equivalence for \quicksort}\label{sec:prove:QS_PE}
In addition to establishing the sortedness property~\eqref{prop:sortedqs} of \quicksort, the functional
correctness of \quicksort{} also
requires proving  the  permutation equivalence
property~\eqref{prop:equal_msets} for \quicksort{}. That is, we prove:
\begin{align}\label{prop:qs:equal_msets}
	\forall x_a, xs_{a} \anddot filter_=(x, xs) &= filter_=(x, quicksort(xs)).
\end{align}

In this respect,
we follow the approach introduced in  Section
\ref{sec:algorithms:perm} to enable first-order reasoning over
permutation equivalence~\eqref{prop:qs:equal_msets}. Namely, we use
$filter_=$ to filter elements $x$ in the lists $xs$ and
$quicksort(xs)$, respectively, and  build the corresponding 
multisets containing as many  $x$ as $x$ occurs in $xs$ and
$quicksort(xs)$. By comparing the resulting multisets, we implicitly
reason about the number of occurrences of $x$ in $xs$ and $quicksort(xs)$, yet, without
the need to explicitly count occurrences of $x$. In summary, we reduce
the task of 
proving~\eqref{prop:qs:equal_msets} to \emph{compositional reasoning} again,
namely to proving following \emph{two
properties given as first-order lemmas} which, by compositionality,
imply~\eqref{prop:qs:equal_msets}: 

\begin{enumerate}[label={{\bf (P\arabic*)}},wide=0em,leftmargin=.5em]
\item\label{P1} List concatenation commutes with $filter_=$, expressed by the lemma: 
\begin{equation}\label{prop:ms:lemma1}
	\begin{split}
		\forall x_a, xs_a,  ys_a \anddot 
		filter_=(x, append(xs, ys)) = append(filter_=(x, xs), filter_=(x, ys)).
	\end{split}
\end{equation}

\item \label{P2} Appending the aggregate of both
  \pv{filter}-operations results in the same multisets as the
  unfiltered list,
  that is, permutation equivalence is invariant over combining complementary
  reduction operations. This property is expressed via: 
\begin{equation} \label{prop:ms:lemma2}
\begin{split}
	\forall x_a, y_a,  xs_a \anddot  
	filter_=(x, xs) = append( & filter_=(x, filter_<(y, xs)), \\
	& filter_=(x, filter_\geq(y, xs))). 
\end{split}
\end{equation}
  \end{enumerate}
  
Similarly as in Section~\ref{sec:prove:sorted}, 
we prove lemmas \ref{P1}--\ref{P2} by
saturation-based reasoning with built-in induction. In particular,
investigating the proof output shows that lemma~\ref{P1} is established using the structural
induction schema~\eqref{eq:list-schema} in saturation,
while the validity of lemma~\ref{P2} is obtained by
applying
the computation induction schema~\eqref{eq:quicksort-schema} in
saturation.

By proving lemmas~\ref{P1}--\ref{P2}, we thus establish validity of 
permutation equivalence ~\eqref{prop:qs:equal_msets} for
\quicksort. Together with the sortedness
property~\eqref{prop:sortedqs} of \quicksort{} proven in
Section~\ref{sec:prove:sorted},
we conclude the functional correctness of \quicksort{} in a fully
automated and compositional manner, using saturation-based theorem proving with built-in
induction and \emph{altogether nine first-order lemmas} in addition to the first-order
semantics~\eqref{eq:div-and-conq} of \quicksort.

\begin{figure}[t]
	\begin{minipage}[]{1\textwidth}%
		\begin{lstlisting}[mathescape]
mergesort :: a' list $\rightarrow$ a' list
mergesort($\nil$) = $\nil$
mergesort(xs) =  merge(mergesort(take((xs$_{length}$ div 2), xs)), mergesort(drop((xs$_{length}$ div 2), xs)))

merge :: a' list $\rightarrow$ a' list $\rightarrow$ a' list
merge($\nil$, ys) = ys
merge(xs, $\nil$) = xs
merge($\cons$(x, xs), $\cons$(y, ys)) = 
	if (x $\leq$ y) {
		$\cons$(x, merge(xs, $\cons$(y, ys)))
	} else {
		$\cons$(y, merge($\cons$(x, xs), ys))
	}	
		\end{lstlisting} \vspace{-1em}
		\caption{ Recursive \mergesort{} over lists of sort $a$. } \label{fig:mergesort}
		
	\end{minipage}\vspace*{-\baselineskip}
\end{figure}

%% file: induction.tex

\section{Compositional Reasoning and Lemma Generalizations}\label{sec:lemmas}%

Establishing the functional correctness of \quicksort{} in
Section~\ref{sec:quicksort} uses nine first-order lemmas that 
express inductive properties over lists
in addition to the first-order
semantics~\eqref{eq:div-and-conq} of \quicksort. While each of these lemmas 
is 
proved by saturation using structural/computation induction schemata,
coming up with proper inductive lemmas remains crucial in reasoning
about inductive data structures, and, so far, dependent on user guidance. 
We thus discuss the intuition on manual proof splitting in Section~\ref{sec:lemmas:split} and generalize our efforts for sorting algorithms in Section~\ref{sec:lemmas:gen}. 

\subsection{Guiding Proof Splitting}\label{sec:lemmas:split}
	Contrary to automated approaches that use inductive annotations to alleviate inductive reasoning, our approach synthesizes the correct induction axioms automatically during saturation runs to prove properties and lemmas correct. However, a manual limitation remains, namely deciding when a lemma is necessary or helpful for the automated reasoner. 
	\newline Splitting the proof into multiple lemmas is necessary
        to guide the prover to find the right terms to apply the
        inductive inferences of
        Section~\ref{sec:semantics:induction}. This is particularly
        the case when input problems, such as sorting algorithms, contain calls to multiple recursive functions - each of which has to be shown to preserve the property that is to be verified. 
        
        In case a proof fails, we investigate the synthesized induction axioms, manually strengthen the property and add any additional assumptions as proof obligations whose validity is in
        turn again verified with the theorem prover and built-in
        induction. That is, we do not simply assume inductive lemmas
        but also provide a formal argument of their validity. 
        
        We illustrate and examine the need
        for proof splitting using lemma \eqref{prop:qs:lemma1}. 

	\begin{example}[Compositional reasoning over sortedness in saturation]\label{rem:comp}
		Consider the following stronger version of lemma~\eqref{prop:qs:lemma1} in the proof of \quicksort:
		\begin{equation} \label{prop:qs:lemma1:strong}
			\begin{array}{l}
				\forall x_a, xs_a, ys_a \anddot 
				 \big( sorted(xs) \wedge sorted(ys)\big)
				\rightarrow
				sorted(append(ys, \cons(x, xs))).
			\end{array}
		\end{equation}
		This formula was automatically be derived by saturation with computation
		induction~\eqref{eq:quicksort-schema} while trying to prove sortedness of the algorithm.
		However, formula~\eqref{prop:qs:lemma1:strong} is not helpful for the proof of \quicksort{} since it is not inductive with regards to the specification and thus cannot be resolved and used during proof search. 
		The prover needs additional information to verify sortedness. Therefore, the assumptions
		$elem_\leq list(x,xs)$ and $list_\leq list(ys,\cons(x, xs))$ are
	  needed in addition to~\eqref{prop:qs:lemma1:strong}, resulting
		in lemma~\eqref{prop:qs:lemma1}. Yet,
		lemma~\eqref{prop:qs:lemma1} from Section~\ref{sec:quicksort} can
		be automatically derived via saturation based on computation
		induction~\eqref{eq:quicksort-schema}. That is, we manually split proof obligations based on missing information in the saturation runs: we
		derive~\eqref{prop:qs:lemma1:strong} from \eqref{eq:quicksort-schema}
		via saturation, strengthen the hypotheses of~\eqref{prop:qs:lemma1:strong} with missing necessary conditions $elem_\leq
			list(x,xs)$ and $list_\leq list(ys,\cons(x, xs))$, and prove their validity via saturation,
		 yielding~\eqref{prop:qs:lemma1}.
	\end{example}

\subsection{Lemma Generalizations for Sorting}\label{sec:lemmas:gen}
The lemmas from Section~\ref{sec:quicksort} represent a number of common proof splits that can be applied to various list sorting tasks. In the following we generalize their structure and apply them to two other sorting algorithms, namely \mergesort{} and \insort.
 \smallskip

\begin{wrapfigure}{r}{0.55\textwidth}
	\begin{lstlisting}[mathescape]
insertsort :: a' list $\rightarrow$ a' list
insertsort($\nil$) = $\nil$
insertsort($\cons$(x, xs)) = insert(x, insertsort(xs))

insert :: a' $\rightarrow$ a' list $\rightarrow$ a' list
insert(x, $\nil$) = $\cons$(x, $\nil$)
insert(x, $\cons$(y, ys)) = 
	if (x $\leq$ y) {
		$\cons$(x, $\cons$(y, ys))
	} else {
		$\cons$(y, insert(x, ys))
	}
	\end{lstlisting} \vspace{-1em}
	\caption{ Recursive algorithm of \insort.}
	\label{fig:insertionsort}
\end{wrapfigure}
\paragraph{Common Patterns of Inductive Lemmas for Sorting Algorithms. }  
\CommentedOut{We first outline our findings with regards to common lemma patterns that emerged from our
investigation of divide-and-conquer
sorting approaches of Figure~\ref{fig:div-and-conq}.
}

Consider the computation induction schema~\eqref{eq:quicksort-schema}.
When using~\eqref{eq:quicksort-schema} for proving the  sortedness~\eqref{prop:sortedqs} and permutation
equivalence~\eqref{prop:qs:equal_msets} of \quicksort{}, the inductive
formula $F$ of~\eqref{eq:quicksort-schema} is, respectively, 
instantiated with the predicates $sorted$ from~\eqref{prop:sortedqs} and
$filter_{=}$ from~\eqref{prop:qs:equal_msets}. 
The base case $F[\nil]$ of
schema~\eqref{eq:quicksort-schema} is then trivially proved by saturation
for both properties~\eqref{prop:sortedqs}
and~\eqref{prop:qs:equal_msets} of \quicksort.

Proving the induction step case of
schema~\eqref{eq:quicksort-schema} is however challenging as it relies
on $partition$-functions which are further used  by $combine$ functions within
the divide-and-conquer patterns of
Figure~\ref{fig:div-and-conq}. Intuitively this means, that proving
the induction step case of
schema~\eqref{eq:quicksort-schema} for the
sortedness~\eqref{prop:sortedqs} and permutation
equivalence~\eqref{prop:qs:equal_msets} properties requires
showing that applying  $combine$ functions over $partition$ functions
preserve sortedness~\eqref{prop:sortedqs} and permutation
equivalence~\eqref{prop:qs:equal_msets},  respectively. For
divide-and-conquer algorithms of Figure~\ref{fig:div-and-conq},
the step case of schema~\eqref{eq:quicksort-schema} requires thus proving
the following lemma: 
\begin{equation}\label{eq:compind:step}
	\begin{pmatrix}\forall
		x_a,ys_a.  \begin{pmatrix} 
			combine\begin{pmatrix}  L[partition_\circ
				(x,ys)],\\
				L[partition_{\circ^{-1}}(x,ys)]\end{pmatrix}\to
			L[\cons(x,ys)])\end{pmatrix}\end{pmatrix}. 
\end{equation} 
We next describe generic instances of lemmas to
be used to prove such step cases and hence functional correctness of sorting algorithms, and exemplify our findings from \quicksort{} by application to recursive versions of \mergesort{} and \insort{} given in Figures~\ref{fig:mergesort} and \ref{fig:insertionsort}, respectively. 
\smallskip

\paragraph{(i) \textit{Combining sorted lists preserves sortedness}. }
For proving the inductive step case~\eqref{eq:compind:step}  of the sortedness property~\eqref{prop:sorted} of sorting algorithms,  we
require the following generic lemma~\eqref{prop:sort:lemma}: 
\begin{equation} \label{prop:sort:lemma}
	\forall xs_a, ys_a \anddot 
	\big( sorted(xs) \wedge sorted(ys)\big)
	\rightarrow
	sorted(combine(xs,ys)), 
\end{equation}
ensuring that combining sorted lists results in a sorted list.
Lemma~\eqref{prop:sort:lemma} is used to 
establish property~\ref{S1} of \quicksort{}, namely used as
lemma~\eqref{prop:qs:lemma1} for proving the preservation of
sortedness under the $append$ function.
\CommentedOut{We recall 
\begin{equation*}
	\begin{array}{ll}
		\forall x_a, xs_a, ys_a \anddot & 
		\big( sorted(xs) \wedge sorted(ys) \wedge 
		elem_\leq list(x,xs)  \wedge \\
		& ~list_\leq list(ys,\cons(x, xs)) \big) \\
		&	\rightarrow
		sorted(append(ys,\cons(x, xs))).
	\end{array}
      \end{equation*}
      }

We showcase the generality of lemma~\eqref{prop:sort:lemma} with \mergesort{} as given in Figure
\ref{fig:mergesort}.
The sortedness property~\eqref{prop:sorted} of \mergesort{} can
be proved by using saturation with lemma~\ref{prop:sort:lemma}; note
that the \pv{merge} function acts as a $combine$
function of~\eqref{prop:sort:lemma}. We thus establish sortedness via the following instance
of~\eqref{prop:sort:lemma}:  
\begin{equation*} \label{prop:merge-sortedness}
  \begin{array}{l}
		\forall xs_a, ys_a \anddot 
		sorted(xs) \wedge sorted(ys)
		\rightarrow
		sorted(merge(xs, ys))
	\end{array}
\end{equation*}
\CommentedOut{
  Note that while we only require Lemma \ref{prop:merge-sortedness} to
  prove the sortedness property of \mergesort, we require a further
  lemma to establish the correctness of  Lemma \ref{prop:ms:lemma1}
  itself due to the complex nature of the \pv{merge}-function. We
  outline the total number of lemmas used to automatically prove our
  selection of sorting algorithms in Section \ref{sec:experiments}.
}

Finally,  lemma~\eqref{prop:sort:lemma} is not purely restricted to
divide-and-conquer routines. When proving the sortedness
property~\eqref{prop:sorted} of the recursive \insort{} algorithm of Figure
\ref{fig:insertionsort},
we apply lemma~\eqref{prop:sort:lemma}on
\pv{insert} to establish preservation of sortedness with saturation: 
\begin{equation*}\label{prop:merge-insert}
	\begin{array}{l}
		\forall x_a, xs_a \anddot 
		sorted(xs) 
		\rightarrow
		sorted(insert(x, xs))
	\end{array}
      \end{equation*}

\paragraph{(ii) \textit{Combining partitions preserves permutation equivalence}.}
Similarly to Section~\ref{sec:prove:QS_PE}, proving permutation equivalence~\eqref{prop:equal_msets}  over
divide-and-conquer sorting algorithms of Figure~\ref{fig:div-and-conq}
is established via the following  two properties:  
\begin{enumerate}[label={{$\bullet$}},wide=0em,leftmargin=0em]
\item As in \ref{P1} for \quicksort, we require  that $combine$ commutes with $filter_=$:
	\begin{equation} \label{prop:permeq:lemma2}
    \begin{split}
		\forall x_a, xs_a, ys_a \anddot 
		filter_=(x,combine(xs,ys)) = combine(filter_=(x,xs),
                                         filter_=(x,ys))
    \end{split}
	\end{equation}  
	\item Similarly to \ref{P2}  for \quicksort{}, we ensure that,
          by 
          combining (complementary) \textit{partition} functions, we 
          preserve~\eqref{prop:equal_msets}. That is, 
	\begin{equation} \label{prop:permeq:lemma1}
    \begin{split}
		\forall x_a, xs_a \anddot 
		filter_=(x,xs) = combine(&filter_=(x,partition_\circ(xs)),\\
                             &filter_=(x,partition_{\circ^-1}(xs)))
    \end{split}
  \end{equation}
Note that lemmas~\ref{P1} and ~\ref{P2} for \quicksort{} are instances
of~\eqref{prop:permeq:lemma2} and~\eqref{prop:permeq:lemma1} respectively, as the $append$ function of
\quicksort{} acts as a $combine$ function and the $filter_<$ and $filter_\geq$
functions are the $partition$ functions of
Figure~\ref{fig:div-and-conq}. 
\smallskip
\end{enumerate}

To
prove the permutation equivalence~\eqref{prop:equal_msets} property of
\mergesort{}, we use
the functions \pv{take} and \pv{drop} as the $partition$
functions of 
lemmas~\eqref{prop:permeq:lemma2}--\eqref{prop:permeq:lemma1}. Doing
so, we 
embed a natural number argument $n$ 
in
lemmas~\eqref{prop:permeq:lemma2}--\eqref{prop:permeq:lemma1}, with
$n$ controlling  how many list elements are \textit{taken} and
\textit{dropped}, respectively, in \mergesort. As such, the
following instances of
lemmas~\eqref{prop:permeq:lemma2}--\eqref{prop:permeq:lemma1} are 
adjusted to \mergesort:
\begin{equation*} \label{prop:merge:ms:lemma2}
  \begin{split}
	\forall x_a, xs_a, ys_a \anddot 
	filter_=(x,merge(xs,ys)) = append(filter_=(x,xs),
                                    filter_=(x,ys)), 
  \end{split}
\end{equation*}
and
\begin{equation*} \label{prop:merge:ms:lemma1}
  \begin{split}
	\forall x_a, n_{\natsort}, xs_a \anddot 
	filter_=(x,xs) = 
		append(
			filter_=(x,take(n, xs)),
			filter_=(x,drop(n, xs))), 
  \end{split}
\end{equation*}
with both lemmas
being proved via saturation, thus establishing permutation equivalence~\eqref{prop:equal_msets}.
%
%

Finally, the generality of
lemmas~\eqref{prop:permeq:lemma2}--\eqref{prop:permeq:lemma1}
naturally pays off when proving the
permutation equivalence property~\eqref{prop:equal_msets} of
\insort{}. Here, we only use a simplified instance
of~\eqref{prop:permeq:lemma2} to prove~\eqref{prop:equal_msets} is
preserved by the auxiliary function \pv{insert}. That is, we use the
following instance of~\eqref{prop:permeq:lemma2}: 
\begin{equation*} \label{prop:insort:ms:lemma1}
\forall x_a, y_a, ys_a \anddot 
filter_=(x,\cons(y, ys)) = filter_=(x,insert(y, ys)), 
\end{equation*}
which is automatically derivable by saturation with computation
induction~\eqref{eq:quicksort-schema}.

\smallskip


We conclude by emphasizing that compositional reasoning in
saturation with computation induction enables us to prove challenging
sorting algorithms in a newly automated manner, replacing the manual
effort in carrying out proofs by induction. 

%% file: experiments.tex

\section{Implementation and Experiments}\label{sec:experiments}

\paragraph{Implementation.} Our work on saturation
with induction in the first-order theory of parameterized lists is
implemented in the first-order prover \vampire~\cite{kovacs2013first}.
In support of parameterization, we extended the SMT-LIB parser of \vampire{} to support parametric
data types from SMT-LIB~\cite{BarFT-SMTLIB} -- version 2.6. In particular,
using the \pv{par} keyword, our parser
interprets \pv{(par (a$_1$ ... a$_n$) ...)} similar to universally quantified blocks where each variable \pv{a$_i$} is a type parameter. 

Appropriating a generic saturation strategy, we adjust
the simplification orderings (LPO)
for efficient equality reasoning/rewrites to our setting.
For example, the precedence of function $quicksort$ is higher
than of symbols $\nil$, $\cons$, $append$, $filter_<$ and
$filter_\geq$, ensuring that $quicksort$
function terms are expanded
to their functional definitions.
\CommentedOut{
most of the problems, with a simplification ordering LPO and precedence that aligns with the
dependencies between the functions used ({\tt --to lpo --sp occurrence}). For example, the symbol $quicksort$ has higher precedence
than the symbols its definition uses, namely $\nil$, $\cons$, $append$, $filter_<$ and $filter_\geq$, ensuring that $quicksort$
function terms introduced are expanded into their
definitions.
}

We further apply recent results of encompassment
demodulation~\cite{DBLP:conf/cade/DuarteK22} to improve equality
reasoning within saturation ({\tt --drc encompass}). We use induction
on data types ({\tt --ind struct}), including complex data type terms
({\tt --indoct on}). \smallskip

\begin{table}[t]
	\small
	\begin{minipage}{.46\linewidth}
		\centering
		\begin{tabular}{l|c|c|c}
			\hline
			\multicolumn{4}{c}{\pv{PermEq}}\\
			\hline
			Benchm. & Pr. & T & Required lemmas\\
			\hline\hline
			\pv{IS-PE} & \checkmark & 0.02 & $\{\pv{IS-PE-L1}\}$\\
			\hline
			\pv{IS-PE-L1} & \checkmark & 0.13 & $\emptyset$ \\
			\hline\hline
			\pv{MS-PE} & \checkmark & 0.06 & $\{\pv{MS-PE-L1},\pv{MS-PE-L2}\}$\\
			\hline
			\pv{MS-PE-L1} & \checkmark* & 0 & - \\
			\hline
			\pv{MS-PE-L2} & \checkmark & 0.03 & $\emptyset$ \\
			\hline
			\pv{MS-PE-L3} & \checkmark & 0.15 & $\emptyset$ \\
			\hline\hline
			\pv{QS-PE} & \checkmark & 0.5 & $\{\pv{QS-PE-L1},\pv{QS-PE-L2}\}$\\
			\hline
			\pv{QS-PE-L1} & \checkmark & 0.05 & $\emptyset$ \\
			\hline
			\pv{QS-PE-L2} & \checkmark & 0.09 & $\emptyset$ \\
			\hline
		\end{tabular}
		\medskip
		\caption{\small Experimental evaluation of proving properties of sorting algorithms, using
			a time limit of 5 minutes on machine with
			AMD Epyc 7502, 2.5 GHz CPU with 1 TB RAM, using 1 core and 16 GB RAM per benchmark. \label{table:experiments} }
	\end{minipage}
\begin{minipage}{.55\linewidth}
		\centering
		\vspace{-1.5em}
		\begin{tabular}{l|c|c|c}
			\hline
			\multicolumn{4}{c}{\pv{Sortedness}}\\
			\hline
			Benchm. & Pr. & T & Required lemmas\\
			\hline\hline
			\pv{IS-S} & \checkmark & 0.01 & $\{\pv{IS-S-L1}\}$ \\
			\hline
			\pv{IS-S-L1} & \checkmark & 8.28 & - \\
			\hline
			\hline
			\pv{MS-S} & \checkmark & 0.08 & $\emptyset$ \\
			\hline
			\pv{MS-S-L1} & \checkmark* & 0 & - \\
			\hline
			\pv{MS-S-L2} & \checkmark & 0.02 & $\emptyset$ \\
			\hline
			\hline
			\pv{QS-S} & \checkmark & 0.09 & $\begin{matrix}\{\pv{QS-S-L1},\pv{QS-S-L2},\\\pv{QS-S-L3}\},\{\pv{QS-S-L1},\\\pv{QS-S-L3},\pv{QS-S-L4}\}\end{matrix}$\\
			\hline
			\pv{QS-S-L1} & \checkmark & 0.27 & $\emptyset$\\
			\hline
			\pv{QS-S-L2} & \checkmark & 0.04 & $\{\pv{QS-S-L4}\}$ \\
			\hline
			\pv{QS-S-L3} & \checkmark & 11.82 & $\{\pv{QS-S-L4},\pv{QS-S-L5}\}$ \\
			\hline
			\pv{QS-S-L4} & \checkmark & 8.28 & $\{\pv{QS-S-L6}\}$ \\
			\hline
			\pv{QS-S-L5} & \checkmark & 0 & $\{\pv{QS-S-L7}\}$ \\
			\hline
			\pv{QS-S-L6} & \checkmark & 0.02 & $\emptyset$\\
			\hline
			\pv{QS-S-L7} & \checkmark & 0.02 & $\emptyset$\\
			\hline
		\end{tabular}
		\vspace{-1em}
	\end{minipage}\\
\pv{IS}, \pv{MS} and \pv{QS} correspond to \insort{},
        \mergesort{} and \quicksort; 
	\pv{S} and \pv{PE} respectively denote sortedness~\eqref{prop:sorted} and
        permutation equivalence~\eqref{prop:equal_msets},  and
      \pv{Li} stands for the $i$-th lemma of the problem. 
\end{table}

  \paragraph{Experimental Evaluation.} We evaluated our approach
  over challenging recursive sorting algorithms taken
  from~\cite{nipkow2021functional}, namely \quicksort,
  \mergesort, and \insort{}. We show that the functional correctness of
  these sorting routines can be 
  verified automatically by means of saturation-based theorem proving
  with induction, as summarized in Table~\ref{table:experiments}.
  
  We divide our experiments according to the specification of sorting algorithms: the first column \pv{PermEq} shows the experiments of all sorting routines w.r.t. permutation equivalence~\eqref{prop:equal_msets}, while \pv{Sortedness} refers to the sortedness~\eqref{prop:sorted} property, together implying
  the functional correctness of the respective sorting algorithm. 
  Here, the inductive lemmas of  Sections~\ref{sec:quicksort}--\ref{sec:lemmas}
  are proven in separate saturation runs of \vampire{} with structural/computation
  induction; these lemmas are then used as input assumptions to
  \vampire{} to prove validity of the respective benchmark.\footnote{Link to experiments upon request due to anonymity.}
%
%
%
A benchmark category \pv{SA-PR[-L$_i$]} indicates that
it belongs to proving the property \pv{PR} for sorting algorithm \pv{SA}, where \pv{PR} is one of \pv{S}
(sortedness~\eqref{prop:sorted}) and \pv{PE} (permutation equivalence~\eqref{prop:equal_msets}) and \pv{SA} is one of \pv{IS}
(\insort), \pv{MS} (\mergesort) and \pv{QS} (\quicksort). Additionally, an optional \pv{Li}
indicates that the benchmark corresponds to the $i$-th lemma for proving the property of
the respective sorting algorithm.

For our experiments, we ran all possible combinations of lemmas
 to determine the minimal lemma dependency for each benchmark. For example, the sortedness property of \quicksort{} (\pv{QS-S})
depends on seven lemmas (see Section~\ref{sec:prove:sorted}),
while the third lemma for this property (\pv{QS-S-L$_3$})
depends on four lemmas (see Section~\ref{sec:prove:QS_PE}).
The second column \pv{Pr.} indicates that \vampire{} solved the benchmark by
using a minimal subset
of needed lemmas given in the fourth column. The third column \pv{T} shows the running time in seconds of the respective saturation run using the first solving strategy identified during portfolio mode. 

To identify the successful configuration, we ran \vampire{} in a portfolio setting for 5 minutes on each benchmark, with strategies
enumerating all combinations of options that we hypothesized to be relevant for these problems. 
In accordance with Table~\ref{table:experiments}, \vampire{} compositionally proves permutation equivalence of \insort{}
and \quicksort{} and sortedness of \mergesort{} and \quicksort{}. Note that sortedness of \mergesort{}
is proven without any lemmas, hence lemma \pv{MS-S-L$_1$} is not needed. 
The lemmas \pv{MS-PE-L$_1$} for the permutation equivalence of \mergesort{}
and \pv{IS-S-L$_1$} for the sortedness of \insort{} could be proven
separately by more tailored search heuristics in \vampire{} (hence $\checkmark*$), but our cluster setup
failed to consistently prove these in the portfolio setting. Further statistics on inductive inferences are provided in Appendix~\ref{appendix:inductions}.
%
%
\CommentedOut{
and permutation equivalence 
      summarizes the number of lemmas needed
in each of our examples. Column \pv{Sorted} and \pv{PermEq} show the
number of lemmas to establish sortedness and permutation equivalence
respectively. From Section \ref{sec:quicksort} it is already obvious
that the sortedness proof of \quicksort{} is by far the most complex
while \mergesort{} could be established by following the ideas
presented in Section \ref{sec:lemmas}. It is easily seen that simple
algorithms that do not require complex recursive call such \insort{}
are very effectively proven by automating computation induction: both
sortedness and permutation equivalence only require one automatically
established lemma respectively.
}
%

%% file: related.tex

\section{Related Work}

While \quicksort{} has been proven correct on multiple occasions, first and foremost in the famous 1971's pen-on-paper proof by Foley and Hoare \cite{foley1971proof}, not many have investigated a fully automated proof of the algorithm. 
A partially automated proof of \quicksort relies on \texttt{Dafny}~\cite{leino2010dafny}, where loop invariants are manually provided~\cite{leino2016quicksort}. While~\cite{leino2016quicksort} claims to prove some of the lemmas/invariants, not all invariants are proved correct (only assumed to be so). Similarly, the \texttt{Why3} framework~\cite{filliatre2013why3} has been leveraged to prove sortedness and permutation equivalence of \mergesort~\cite{levy2014simple} over parameterized lists and arrays. These proofs also rely on manual proof splitting with the additional overhead of choosing the underlying prover for each subgoal as \texttt{Why3} is interfaced with both automated and interactive provers. 

The work of \cite{walther2004verification} reports on the verification of functional implementations of multiple sorting algorithms with \texttt{VeriFun}~\cite{walther2003verifun}. Specifically, the correctness of the sortedness property of \quicksort{} is established with the help of 13 auxiliary lemmas while also establishing the permutation property of \mergesort{} by comparing the number of elements, thus requiring additional arithmetic reasoning. In contrast, our proofs involve less auxiliary lemmas, avoid the overhead of arithmetic theories through our formalization of the permutation property over set equivalence and prove functional implementations with arbitrary sorts permitting a linear order. 

The approach of~\cite{safari2020generic} establishes the correctness of permutation equivalence for multiple sorting algorithms based on separation logic through inductive lemmas. 
However, \cite{safari2020generic} does not address the correctness proofs of the sortedness property.
We instead automate the correctness proofs of sorting algorithms via compositional first-order reasoning in the theory of parameterized lists. 

Verifying functional correctness of sorting routines has also been explored in the abstract interpretation and model-checking communities, by investigating array-manipulating programs \cite{DBLP:conf/popl/GulwaniMT08,DBLP:conf/cav/JhalaM07}. 
In \cite{DBLP:conf/popl/GulwaniMT08}, the authors automatically generate loop invariants for standard sorting algorithms of arrays of fixed length; the framework is, however, restricted solely to inner loops and does not handle recursive functions. 
Further, in \cite{DBLP:conf/cav/JhalaM07} a priori given invariants/interpolants are used in the verification process. Unlike these techniques, we do not rely on a user-provided inductive invariant, nor are we restricted to inner loops. 

There are naturally many examples of proofs of sorting algorithms using interactive theorem proving, see e.g. \cite{jian2017insertion,lammich2020efficient}. The work of~\cite{jian2017insertion} establishes correctness of insertion sort. Similarly, the setting of~\cite{lammich2020efficient}  proves variations of \pv{Introsort} and \pv{Pdqsort} -- both using  \texttt{Isabelle/HOL}~\cite{DBLP:conf/tphol/WenzelPN08}. However, interactive provers rely on user guidance to provide induction schemes, a burden that we eliminate in our approach.

A verified a real-world implementation of \quicksort{} is given in~\cite{beckert2017proving}, Here,  Java's inbuilt dual pivot \quicksort{} class is verified with the semi-automatic \texttt{KeY} prover~\cite{ahrendt2005key}. Additionally, the \texttt{KeY} prover has also been leveraged to analyze industrial implementations of \texttt{Radixsort} and \texttt{Countingsort} \cite{de2016verification}. By relying on inductive method annotations,  such as loop invariants or method contracts,  and asking the user to guide the proof rule application during the verification process, the work faces similar limitations as the ones using \texttt{Dafny}, \texttt{Why3} or interactive proving. While we manually split our proofs into multiple steps, our lemmas are proved automatically thanks to saturation-based theorem proving with structural/computation induction. As such, we
do not require guidance on rule application or inductive annotations. 


When it comes to the landscape of automated saturation-based
reasoning, we are not aware of other techniques enabling the fully
automated verification of such sorting routines, with or without compositional reasoning. 

%% file: conclusion.tex

\section{Conclusion and Future Work}
We present an integrated formal approach to establish program correctness over recursive programs based on saturation-based theorem proving. 
We automatically prove recursive sorting algorithms, particularly the \quicksort{} algorithm, by formalizing program semantics in the first-order theory of parameterized lists.
Doing so, we expressed the common properties of sortedness and permutation equivalence in an efficient way for first-order theorem proving. 
By leveraging common structures of divide-and-conquer sorting algorithms, we advocate compositional first-order reasoning with built-in structural/computation induction.

\textcolor{black}{\newline We believe the implications of our work are twofold. First, integrating inductive reasoning in automated theorem proving to prove (sub)goals during interactive theorem proving can significantly alleviate the use of proof obligations to be shown manually,  since automated theorem proving from our work can synthesize induction hypotheses to verify these conditions. Second,  finding reasonable strategies to automatically split proof obligations on input problems can tremendously enhance the degree of automation in proofs that require heavy inductive reasoning. We hope that our work opens up future directions in combining interactive and automated reasoning by further decreasing  the amount of manual work in proof splitting, allowing superposition frameworks to be better applicable to a wider range of recursive algorithms.}
%
Proving  further recursive sorting/search algorithms in future work, with improved compositionality, is therefore an interesting challenge to investigate. 


%
%

%% file: appendix.tex
\appendix
\section{Generated Inductive Inference during Proof Search}\label{appendix:inductions}
For all conjectures and lemmas that were proved in portfolio mode, we summarized the applications of inductive inferences with structural and computation induction schemata in Table~\ref{table:inductions}. Specifically, 
Table~\ref{table:inductions} compares the number of inductive inferences performed during proof search (column \pv{IndProofSearch}) with the number of used inductive inferences as part of each benchmark's proof (column \pv{IndProof}). 
While most safety properties and lemmas required less than 50 inductive inferences, thereby using mostly one or two of them in the proof, some lemma proofs exceeded this by far. Most notably \pv{IS-S-L1} and \pv{QS-S-L1}, \insort's and \quicksort's first lemma respectively, depended on many more inductive inferences until the right axiom was found. 
Such statistics point to areas where the prover still has room to be finetuned for software verification and quality assurance purposes, here especially towards establishing correctness of functional programs. 

\begin{table}
	\centering
	\small
	\caption{Structural Induction Applications in Proof Search and Proof.\label{table:inductions}}
	\begin{tblr}{
			column{2} = {c},
			column{3} = {c},
			hlines,
			vlines,
		}
		\textbf{Benchmark} & \textbf{\pv{IndProofSearch}} & \textbf{\pv{IndProof}} \\
		\pv{IS-S}               & 4                         & 1                   \\
		\pv{IS-S-L1}           & 339                       & 2                   \\
		\pv{IS-PE}              & 5                         & 1                   \\
		\pv{IS-PE-L1}           & 34                        & 1                   \\
		\pv{MS-S}               & 8                         & 1                   \\
		\pv{MS-S-L2}            & 22                        & 1                   \\
		\pv{MS-PE}              & 14                        & 1                   \\
		\pv{MS-PE-L2}           & 16                        & 1                   \\
		\pv{MS-PE-L3}           & 136                       & 3                   \\
		\pv{QS-S}               & 10                        & 2                   \\
		\pv{QS-S-L1}            & 510                       & 2                   \\
		\pv{QS-S-L2}            & 9                         & 1                   \\
		\pv{QS-S-L3}            & 130                       & 2                   \\
		\pv{QS-S-L4}            & 183                       & 3                   \\
		\pv{QS-S-L5}            & 0                         & 0                   \\
		\pv{QS-S-L6}            & 26                        & 1                   \\
		\pv{QS-S-L7}            & 16                        & 2                   \\
		\pv{QS-PE }             & 12                        & 1                   \\
		\pv{QS-PE-L1 }          & 10                        & 1                   \\
		\pv{QS-PE-L2 }          & 42                        & 4                   
	\end{tblr}
\end{table}